\documentclass[showpacs,pre,amssymb,amsmath,preprint]{revtex4}

\usepackage{psfrag}
\usepackage{amsmath}
\usepackage{amssymb}
\usepackage{textcomp}		
\usepackage{array}		
\usepackage{multirow}		
\usepackage{rotating}		
\def \dd{\hbox{d}}

\begin{document}

\title{APPROACH TO A STATIONARY STATE IN AN EXTERNAL FIELD}
\author{A.Alastuey*   J.Piasecki**}
\affiliation{* ENS Lyon, CNRS, France\\ **Institute of Theoretical Physics, University of Warsaw, Ho\.za 69, 00 681 Warsaw, Poland}
\date{\today}
\begin{abstract}

We study relaxation towards a stationary out of equilibrium state by analyzing a one-dimensional 
stochastic process followed by a particle accelerated by an external field and propagating 
through a thermal bath.
The effect of collisions is described within Boltzmann's kinetic theory. 
We present analytical solutions for the Maxwell gas and for the very hard particle model. 
The exponentially fast relaxation of the velocity distribution toward the stationary form is demonstrated.
In the reference frame moving with constant drift velocity the hydrodynamic diffusive mode is shown to govern
the distribution in the position space.
We show that the exact value of the diffusion coefficient for any value of the field 
is correctly predicted by the Green-Kubo autocorrelation formula generalized to the stationary state.

\end{abstract}

\maketitle

\section{Introduction}

The present paper is devoted to the 
the study of a stochastic process followed by a particle  
moving through a scattering thermal bath 
while accelerated by an external field. 
The field prevents the particle from acquiring the Maxwell distribution of the bath. 
Our aim here is not only to establish the precise form of the stationary velocity distribution, 
as it was e.g. the case in the analysis presented in \cite{GP86}, but  also to answer 
the physically relevant question of the dynamics of approach towards the long-time 
asymptotic state.  The evolution of the distribution in 
position space will be thus also discussed. 

\bigskip

We consider a one-dimensional dynamics described by  
the Boltzmann kinetic equation
\begin{equation}
\label{I1}
\left( \frac{\partial}{\partial t}+v\frac{\partial}{\partial r} + 
a\frac{\partial}{\partial v} \right)f(r,v;t)=v_{\text{\tiny int}}^{1-\gamma}
\rho \int \dd w |v-w|^{\gamma}[\, f(r,w;t)\,\phi(v)-f(r,v;t)\,\phi(w)\,]
\end{equation}
Here $f(r,v;t)$ is the probability density for finding 
the propagating particle at point $r$ with velocity $v$ at time $t$. The 
thermal bath particles are not coupled to the external field. Before binary encounters with the accelerated particle they are assumed to be in an equilibrium state with uniform temperature $T$ and density $\rho$
\begin{equation}
\label{I2}
\rho \,\phi(v) = \rho \sqrt{\frac{m}{2\pi k_{B}T}}\exp\left(-\frac{mv^{2}}{2k_{B}T} \right)= 
 \frac{\rho}{v_{\text{\tiny th}}\sqrt{2\pi}}\exp\left[-\frac{1}{2}\left(\frac{v}{v_{\text{\tiny th}}} \right)^{2}\right]
\end{equation}
Here $\phi(v)$ is the Maxwell distribution, and
\begin{equation}
\label{I3}
v_{\text{\tiny th}} = \sqrt{\frac{k_{B}T}{m}} \,. 
\end{equation}
denotes the corresponding thermal velocity.
The differential operator on the left-hand side of (\ref{I1}) 
generates motion with a constant acceleration $a$.
The accelerated motion is permanently perturbed 
by instantaneous exchanges of velocities with thermalized bath particles. 
This is modeled by the Boltzmann collision term on the right hand side of equation (\ref{I1}),
which accounts for elastic encounters between equal mass particles. The collision frequency 
depends therein on the absolute relative velocity
$|v-w|$ through a simple power law with exponent $\gamma$. Finally $v_{\text{\tiny int}}$ is some characteristic velocity of the underlying interparticle interaction. 

\bigskip

In the case of hard rods ($\gamma = 1$) the factor $|v-w|$ 
is the main source of difficulties in the attemps to rigorously determine 
the evolution of $f(r,v;t)$, since it prevents the effective use of 
Laplace and Fourier transformations.  
It was thus quite remarkable that a stationary velocity distribution could be 
analytically determined in that case, leading in particular to an explicit 
expression for the current at any value of the external acceleration \cite{GP86}. In that case, kinetic equation (\ref{I1}) 
has been solved exactly 
only at zero temperature where $\phi (v)|_{T=0} = \delta(v)$ \cite{JP83}.
Also, when $\phi(v)$ is replaced by the distribution 
$[\delta(v-v_{0})+\delta(v+v_{0})]/2$ with a discrete velocity spectrum $\pm v_{0}$, an explicit analytic solution 
has been derived and analyzed in \cite{JP1986} and \cite{JPRS2006}.
The physically relevant conclusions from those works can be summarized as follows
\begin{itemize}
\item[(i)] the approach to the asymptotic stationary velocity distribution is exponentially fast
\item[(ii)] in the reference system moving with average velocity,
the hydrodynamic diffusion mode governs the spreading of the distribution in position space
\item[(iii)] the Green-Kubo autocorrelation formula for the diffusion coefficient applies in the non-equilibrium steady state
\end{itemize}

Our aim is to show that the general features (i)-(iii) persist when $\phi(v)$ 
is the Maxwell distribution with temperature $T>0$. However, in 
the present study, we restrict the analysis 
to cases $\gamma =0$ and $\gamma =2$, 
which are much simpler than the hard-rod one. Indeed, it turns out that
the Fourier-Laplace transformation can then be effectively used to solve 
the initial value problem for equation (\ref{I1}). 
The simplifications occuring when $\gamma =0$ or $\gamma =2$ 
have been already exploited in other studies: 
for recent applications to granular fluids, see e.g. 
\cite{BCG2000}-\cite{MP2007} and references quoted therein.

\bigskip

In terms of dimensionless variables
\begin{equation}
\label{I4}
w = v/v_{\text{\tiny th}}, \;\;\;\; 
x=r \, \rho \left(v_{\text{\tiny th}}/v_{\text{\tiny int}}\right)^{\gamma-1}, 
\;\;\;\; \tau = t \,\rho \, v_{\text{\tiny th}} 
\left(v_{\text{\tiny th}}/v_{\text{\tiny int}}\right)^{\gamma-1}\, ,
\end{equation}
the kinetic equation (\ref{I1}) takes the form
\begin{equation}
\left( \frac{\partial}{\partial \tau}+w\frac{\partial}{\partial x} + 
\epsilon\frac{\partial}{\partial w} \right)F(x,w;\tau) =
\int \dd u |w-u|^{\gamma} [F(x,u;\tau)\Phi(w)-F(x,w;\tau)\Phi(u)] \, ,
\label{I5}
\end{equation}
where $\Phi(w)$ is the dimensionless normalized gaussian
\begin{equation}
\label{I7}
\Phi(w)= \frac{1}{\sqrt{2\pi}}e^{-w^{2}/2} \, ,
\end{equation}
and $\epsilon$ is the dimensionless parameter 
\begin{equation}
\label{I6}
\epsilon = \left(v_{\text{\tiny th}}/v_{\text{\tiny int}}\right)^{1-\gamma} \,
\frac{am\rho^{-1}}{ k_{B}T}\
\end{equation}
proportional to the ratio between the energy $am\rho^{-1}$ provided 
to the particle on a mean free path, 
and thermal energy $k_{B}T$. That parameter
can thus be looked upon as a measure of the strength of the field.
Integration of (\ref{I5}) over the position space yields 
the kinetic equation for the velocity distribution 
\[ G(w;\tau)=\int \dd x F(x,w;\tau) \; ,\]
which reads
\begin{equation}
\left( \frac{\partial}{\partial \tau} + \epsilon\frac{\partial}{\partial w} \right)G(w;\tau) =
\int \dd u |w-u|^{\gamma} [G(u;\tau)\Phi(w)-G(w;\tau)\Phi(u)] \, .
\label{I8}
\end{equation}

\bigskip

The paper is organized as follows. In Section II, we 
consider the so-called Maxwell gas ($\gamma =0$). The explicit solution of the kinetic 
equation (\ref{I5}) enables a thorough discussion of the approach to the stationary state, together with a study of the structure of the stationary velocity distribution. In Section III, we proceed to 
a similar analysis for the very hard particle model ($\gamma =2$). Section IV contains conclusions. Some calculations have been relegated to Appendices. 

\section{The Maxwell gas}

We consider here the simple version $\gamma =0$ of equation (\ref{I5}). One usually then
refers to the Maxwell gas dynamics, in which the collision frequency 
does not depend on the speed of approach (see e.g. \cite{UFM63}).
This case can be viewed upon as a very crude approximation to the 
hard rod dynamics ($\gamma=1$) obtained by replacing the relative speed $|v-c|$ of colliding particles 
by constant thermal velocity $v_{\text{\tiny th}}$, while $v_{\text{\tiny int}}$ is identified with $v_{\text{\tiny th}}$.
Here, kinetic equation (\ref{I5}) takes the form
\begin{eqnarray}
\left( \frac{\partial}{\partial \tau}+w\frac{\partial}{\partial x} + 
\epsilon\frac{\partial}{\partial w} \right)F(x,w;\tau) & =
 & \int \dd u [F(x,u;\tau)\Phi(w)-F(x,w;\tau)\Phi(u)] \nonumber \\
& = &  M_{0}(x;\tau)\Phi(w) - F(x,w;\tau) \label{II1}
\end{eqnarray}
where $M_{0}(x;\tau )$ denotes the zeroth moment
\begin{equation} 
\label{II2}
 M_{0}(x;\tau) = \int \dd u F(x,u;\tau) \; .
\end{equation}

\bigskip

Equation (\ref{II1}) can be conveniently rewritten as an integral equation 
\begin{multline}
F(x,w;\tau) = e^{-\tau}F(x-w\tau+\epsilon\tau^{2}/2, w-\epsilon\tau;0) \\
+ \int_{0}^{\tau}d\eta e^{-\eta}\Phi(w-\epsilon\eta)M_{0}(x-w\eta+\epsilon\eta^{2}/2; \tau-\eta) \, ,
\label{II3} 
\end{multline}
with an explicit dependence on the initial condition $F(x,w;0)$.
Integration of equation (\ref{II3}) over $x$ yields 
\begin{equation}
\label{II4}
G(w;\tau) = \int \dd x F(x,w;\tau) = e^{-\tau}G_{\text{\tiny in}}( w-\epsilon\tau) + 
N_{0} \int_{0}^{\tau} \dd \eta e^{-\eta}\Phi(w-\epsilon\eta) \, ,
\end{equation}
where $G_{\text{\tiny in}}(w) = G(w;0)$ is the initial condition, and $N_{0}=\int \dd w \int \dd x F(x,w;\tau) = \int \dd w G(w;\tau)$ is the conserved normalization factor.

\subsection{Stationary solution and relaxation of the velocity distribution}

Putting   $N_{0}=1$ in formula (\ref{II4}) yields the evolution law for the normalized velocity distribution
\begin{equation}
\label{IIG}
G(w;\tau) = \int \dd x F(x,w;\tau) = e^{-\tau}G_{\text{\tiny in}}( w-\epsilon\tau) + 
\int_{0}^{\tau} \dd \eta e^{-\eta}\Phi(w-\epsilon\eta) \, ,
\end{equation}
The first term on the right hand side of (\ref{IIG}) describes the 
decaying memory of the initial distribution : $G_{\text{\tiny in}}(w)$ 
propagates in the direction of the field with constant velocity $\epsilon$, 
while its amplitude is exponentially damped. Clearly, for times $\tau \gg 1$ 
that term can be neglected.

\bigskip

\begin{figure}
\includegraphics[width=0.9\textwidth]{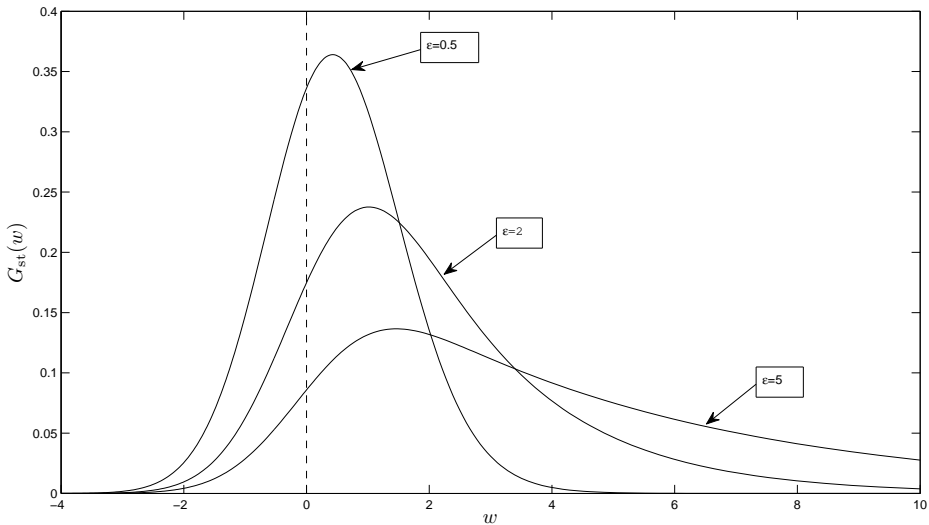}
\caption{\label{AP09a} Stationary velocity distribution 
$G_{\text{\tiny st}}(w)$ for three values of $\epsilon$.}
\end{figure}

The second term in formula (\ref{IIG}) describes the approach 
to the asymptotic stationary distribution 
\begin{eqnarray}
\label{II5}
G_{\text{\tiny st}}(w) = G(w;\infty) & = & \int_{0}^{\infty} \dd \eta 
\; e^{-\eta}\;\Phi(w-\epsilon\eta) \nonumber \\
&=& \frac{1}{2\epsilon}\exp{\left(\frac{1}{2\epsilon^{2}}-\frac{w}{\epsilon} \right) } 
\left( 1+\text{Erf}\left(\frac{w\epsilon-1}
{\epsilon\sqrt{2}}\right)\right)\, ,
\end{eqnarray}
where 
\[ \text{Erf}(\xi)=\frac{2}{\sqrt{\pi}}\int_0^\xi \dd u \; \exp(-u^2) \] 
is the familiar error function.
It is interesting to compare the decay-law of 
$G_{\text{\tiny st}}(w)$ at large velocities, to that 
corresponding to the case of hard-rod collisions. Using expression (\ref{II5})
we find the asymptotic formula
\begin{equation}
\label{II16}
G_{\text{\tiny st}}(w) \sim \frac{1}{\epsilon}
\exp{\left(\frac{1}{2\epsilon^{2}}-\frac{w}{\epsilon} \right) }
\end{equation}
when $w \to +\infty$. In contradistinction to the hard-rod case governed 
by an $\epsilon$-dependent gaussian law (see \cite{GP86}) 
we find here a purely exponential decay. The thermal bath is unable 
to impose via collisions its own gaussian decay because of insufficient collision frequency.
The replacement of the relative speed in the Boltzmann 
collision operator by thermal velocity implies thus qualitative changes in
the shape of the stationary velocity distribution. The plot of 
$G_{\text{\tiny st}}(w)$ for different values of $\epsilon$ is shown in Fig.~\ref{AP09a}.

\bigskip

Basic properties (i)-(iii) discussed in the 
Introduction turn out to be valid. Indeed, the inequality
\begin{equation}
\label{II6}
G_{\text{\tiny st}}(w) -  \int_{0}^{\tau}\dd \eta 
\; e^{-\eta}\; \Phi(w-\epsilon\eta) 
=\int_{\tau}^{\infty}\dd \; \eta e^{-\eta}\; \Phi(w-\epsilon\eta) 
< \frac{e^{-\tau}}{\epsilon} 
\end{equation}
displays an uniform exponentially fast approach towards the stationary state.
In particular, using formula (\ref{IIG}), we find that the average velocity 
$<w>(\tau)$ approaches the asymptotic value
\begin{equation}
\label{II7}
<w>_{\text{\tiny st}} = \epsilon
\end{equation}
according to 
\begin{equation}
\label{II8}
<w>(\tau) = \int \dd w \, w\,  G(w;\tau) = \epsilon + 
e^{-\tau}[ <w>_{\text{\tiny in}} - \epsilon ]
\end{equation}
We encounter here an exceptional situation 
where the linear response is exact for any value of 
the external field. 

\bigskip

Equation (\ref{II4}) with $N_{0}$ put equal to zero can be used for the evaluation of the time-displaced velocity autocorrelation 
function 
\begin{equation}
\label{II9}
\Gamma(\tau) = <[ w(\tau) - <w>_{\text{\tiny st}} ]
[w(0) - <w>_{\text{\tiny st}} ]>_{\text{\tiny st}} \, .
\end{equation}
where $<...>_{\text{\tiny st}}$ denotes the average over 
stationary state (\ref{II5}). The calculation presented in Appendix~\ref{B}
provides the formula
\begin{equation}
\label{II10}
\Gamma(\tau) = e^{-\tau} [ 1 + \epsilon^{2} ] \, ,
\end{equation}
which yields a remarkably simple field dependence of the diffusion coefficient
\begin{equation}
\label{II11}
D(\epsilon) = \int_{0}^{\infty} \dd \tau \; \Gamma(\tau) = 1 + \epsilon^{2} \, .
\end{equation}

\subsection{Relaxation of density: appearence of a hydrodynamic mode}

Let us turn now to the analysis of the evolution of the normalized density $n(x;\tau)=M_{0}(x;\tau)$ 
in position space. It turns out that one can solve 
the complete integral equation (\ref{II3}) by applying 
to both sides Fourier and Laplace transformations. If we set
\begin{equation}
\label{II12bis}
\tilde{F}(k,w;z) = \int_0^{\infty} \dd \tau\, e^{-z\tau} 
\int \dd x \, e^{-ikx}\, F(x,w;\tau) \, , 
\end{equation}
we find
\begin{multline}
\label{II12}
\tilde{F}(k,w;z) = \int_{0}^{\infty}\dd \tau \; {\rm exp}
\left[ -ik\left( w\tau - \epsilon\frac{\tau^{2}}{2} \right) - 
(z+1)\tau  \right] \\
\left\lbrace \hat{F}_{\text{\tiny in}}(k,w-\epsilon\tau) 
+ \tilde{n}(k;z) \Phi(w-\epsilon\tau) \right\rbrace \; ,
\end{multline}
where $\tilde{n}(k;z)$ is the Fourier-Laplace transform of $n(x;\tau)$, and 
\[ \hat{F}_{\text{\tiny in}}(k,w)= \int \dd x \, e^{-ikx}\, F(x,w;0) \] 
denotes the spatial Fourier transform of the initial condition.
Equation (\ref{II12}) when integrated over the velocity space yields the formula
\begin{equation}
\label{II13}
\tilde{n}(k;z) =\frac{1}{\zeta(k;z)}\int \dd w  \int_{0}^{\infty}\dd \tau \; {\rm exp}
\left[ -ik\left( w\tau - \epsilon\frac{\tau^{2}}{2} \right) - (z+1)\tau  \right]
\hat{F}_{\text{\tiny in}}(k,w-\epsilon\tau)   
\end{equation}
with
\begin{equation}
\label{II14}
\zeta(k;z) = 1 - \int_{0}^{\infty}\dd \tau \; {\rm exp}
\left[ - (z+1)\tau - (ik\epsilon + k^{2})\frac{\tau^{2}}{2} \right]  \, .
\end{equation}
The insertion of (\ref{II13}) into (\ref{II12}) provides 
a complete solution for $\tilde{F}(k,w;z)$ corresponding to a given initial condition.

\bigskip

Formula (\ref{II13}) shows that the time-dependence of the 
spatial distribution is defined by roots of the function 
$\zeta(k;z)$. In order to find the long-time hydrodynamic mode 
$z_{\rm{hy}}(k)$, we have to look for the root of $\zeta(k;z)$ 
which approaches $0$ when $k\to 0$. If we assume the asymptotic form
\[ z_{\rm{hy}}(k) = c_{1}k + c_{2}k^{2} +o(k^2) \;\;\; \text{when} 
\;\;\; k\to 0\; , \]
we find a unique self-consistent solution to  equation 
$\zeta(k;z)=0$ of the form
\begin{equation}
\label{II15}
z_{\rm{hy}}(k)= -i\epsilon k - (1+\epsilon^{2})k^{2} + o(k^2)  = 
-i\epsilon k - D(\epsilon) k^{2} + o(k^2) \, .
\end{equation}
It has the structure of a propagating diffusive mode. 
It is important to note that
the diffusion coefficient $D(\epsilon)$ equals $(1+\epsilon^{2})$ 
in accordance with the Green-Kubo result (\ref{II11}). 
We thus see that, in the reference system moving with 
constant velocity $\epsilon$,
a classical diffusion process takes place in position space. 

\bigskip

It has been argued in the literature that, in general, 
$z_{\rm{hy}}(k)$ is not an analytic function of $k$ at $k=0$ 
(see \textsl{e.g.} Ref.~\cite{ED75}). Here, that question can be 
precisely investigated as follows. According to the integral
expression (\ref{II14}) of $\zeta(k;z)$, the hydrodynamic mode is 
a function of $\xi=ik\epsilon + k^2$. By combining differentiations 
with respect to $\xi$ under the integral sign with 
integration by parts, we find that $z_{\rm{hy}}(\xi)$ satisfies the 
second order differential equation 
\begin{equation}
\label{IIhyddiff}
\xi \frac{\dd^2 z_{\rm{hy}}^2}{\dd \xi^2} = 1+ 
\frac{\dd z_{\rm{hy}}}{\dd \xi} \; .
\end{equation} 
Then, since $z_{\rm{hy}}(0)=0$, we find that $z_{\rm{hy}}(\xi)$ can be 
formally represented by an infinite entire series in $\xi$, 
\begin{equation}
\label{IIhydTaylor}
z_{\rm{hy}}(\xi) = \sum_{n=1}^{\infty} c_n \xi^n \; , 
\end{equation}
with $c_1=-1$, $c_2=1$ and 
\[ |c_{n+1}| \geq 2^{n-1} \; n! \;\;\; \text{for} \;\;\; n \geq 2 \; .\] 
Thus, the radius of convergence of Taylor series (\ref{IIhydTaylor}) 
is zero, so $\xi=0$ is a singular point of function $z_{\rm{hy}}(\xi)$, as 
well as $k=0$ is a singular point of function $z_{\rm{hy}}(k)$. The nature 
of that singularity can be found by rewriting the root equation defining 
$z_{\rm{hy}}(\xi)$ as the implicit equation
\begin{equation}
\label{IIhydimp}
1-\text{Erf}\left(\frac{z_{\rm{hy}}+1}{\sqrt{2\xi}}\right) = 
\sqrt{\frac{2\xi}{\pi}} \; 
\exp\left(-\frac{(z_{\rm{hy}}+1)^2}{2\xi}\right) \; .
\end{equation}
The introduction of function $\sqrt{\xi}$ requires to define cut-lines 
ending at points $k=0$ and $k=-i\epsilon$ which are the two roots of 
equation $\xi(k)=0$. Since the integral in the r.h.s. of 
expression (\ref{II14}) diverges for $k$ imaginary of the form 
$k=iq$ with $q > 0$ or $q < -\epsilon$, it is natural to define 
such cut-lines as $[i0, i\infty[$ and $]-i\infty, -i\epsilon]$. The 
corresponding choice of determination for $\sqrt{\xi}$ is defined by 
$\sqrt{\xi(k^+)}= i\sqrt{q\epsilon+q^2}$ for $k^+=0^+ + iq$ with $q>0$, 
where $\sqrt{q\epsilon+q^2}$ is the usual real positive square root of the 
real positive number $(q\epsilon+q^2)$. Notice that, when 
complex variable $k$ makes a complete tour around point $k=0$ starting 
from $k^+=0^+ + iq$ on one side of the cut-line 
and ending at $k^-=0^- + iq$ on 
the other side (with vanishing difference $k^+-k^-$),  
$\sqrt{\xi(k)}$ changes sign from $\sqrt{\xi^+}$ to 
$\sqrt{\xi^-}=-\sqrt{\xi^+}$ with obvious notations. As shown by adding 
both implicit equations (\ref{IIhydimp}) for 
$k^+$ and $k^-$ respectively, $z_{\rm{hy}}^+$ does not reduce to 
$z_{\rm{hy}}^-$. The difference $(z_{\rm{hy}}^+-z_{\rm{hy}}^-)$ is of order 
$\exp(-1/(2|k|\epsilon))$, so $k=0$ is an essential singularity.

\section{Very hard particles}

Another interesting case is that of the so-called very hard particle model, 
where the collision frequency is proportional to the 
kinetic energy of the relative motion of the colliding pair. 
The corresponding 
exponent in the collision term of the Boltzmann equation (\ref{I1}) 
is now $\gamma=2$. This allows us
to simplify the resolution of the kinetic equation. Owing to this fact,
the very hard particle model, similarly to the Maxwell gas, has been studied in 
numerous works (see e.g. \cite{MHE1984}-\cite{CDT2005}, and references given therein).

\bigskip

Using dimensionless variables (\ref{I4}), we thus write 
the kinetic equation as
\begin{equation}
\left( \frac{\partial}{\partial \tau}+w\frac{\partial}{\partial x} + \epsilon\frac{\partial}{\partial w} \right)F(x,w;\tau) =
 \int \dd u |w-u|^{2} [F(x,u;\tau)\Phi(w)-F(x,w;\tau)\Phi(u)] 
\label{III1}
\end{equation}
\[ = [ w^2 M_{0}(x;\tau)-2w M_{1}(x;\tau)+M_{2}(x;\tau)]\Phi(w)-(w^2+1)F(x,w;\tau)\]
where the moments $M_{j}(x;\tau)$ ($j=1,2,...$) are defined by
\begin{equation}
\label{IIIM}
M_{j}(x;\tau) = \int \\d w w^j F(x,w;\tau) \; .
\end{equation}
The evolution equation of the velocity distribution $G(w;\tau)$ becomes
\begin{equation}
\label{III2}
\left( \frac{\partial}{\partial \tau}+\epsilon\frac{\partial}{\partial w} \right)G(w;\tau) =
[ N_{2}(\tau)-2wN_{1}(\tau)+w^{2}N_{0}]\Phi(w)-(w^{2}+1)G(w;\tau) \, ,
\end{equation}
with the integrated moments
\begin{equation}
\label{IIIN}
N_{j}(\tau)=\int \dd x\,M_{j}(x;\tau), \;\; j=0,1,2 \; .
\end{equation}
Notice that the integrated zeroth moment does not depend on time since the evolution 
conserves the initial normalization condition 
\[N_{0}(\tau)=\int \dd w \int \dd x F(x,w;\tau) = N_{0}\; .\]
Hence, when $F(x,w;\tau)$ is a normalized probability density $N_{0}(\tau)=N_{0}=1$.

\bigskip

The simplification related to the 
choice $\gamma=2$, and more generally when $\gamma$ is an even integer, 
concerns the collision term in the
general kinetic equation (\ref{I1}) which can be expressed in such cases in
terms of a finite number of moments of the distribution function. 
The resolution of that equation becomes then straightforward within 
standard methods (see Appendix~\ref{A}).

\subsection{Laplace transform of the velocity distribution}

The expression for the Laplace transform of the normalized velocity distribution follows directly from the general formula 
(\ref{C3}) derived in Appendix~\ref{A} by putting $k=0$, and choosing
$\tilde{M}_{0}(0,z)= \tilde{N}_{0}(z) = 1/z$. 
Within definition 
\begin{equation}
\label{III7}
S(w;z) =(z+1) w + \frac{w^3}{3} \,  
\end{equation}
for the function $S(k,w;z)$ evaluated at $k=0$ (see definition (\ref{S})),
we find
\begin{multline}
\label{III6}
\epsilon \tilde{G}(w;z) =  \frac{\epsilon}{z}\Phi(w) + 
\int_{-\infty}^{w}\dd u \exp \{ [S(u;z)-S(w;z)]/\epsilon \}\; \{ 
G_{\text{\tiny in}}(u)  \\
+ [\tilde{N}_{2}(z)-2u\tilde{N}_{1}(z)+ \frac{(\epsilon u -z-1)}{z}] 
\Phi(u) \} \; . 
\end{multline}
\bigskip
The two functions $\tilde{N}_{1}(z)$ and $ \tilde{N}_{2}(z)$ 
satisfy the system of equations 
\begin{eqnarray}
\label{III8}
0 & = & A^{\text{\tiny (in)}}_{00}(0;z) + [ \tilde{N}_{2}(z) - (z+1)/z]A_{00}(0;z)+[\epsilon/z -2\tilde{N}_{1}(z)]A_{01}(0;z) \nonumber \\
\epsilon \tilde{N}_{1}(z) & = &  A^{\text{\tiny (in)}}_{10}(0;z) + [ \tilde{N}_{2}(z) - (z+1)/z]A_{10}(0;z)+[\epsilon/z-2\tilde{N}_{1}(z)]A_{11}(0;z)
\end{eqnarray}
which is identical to (\ref{C7}) taken at $k=0$, while  
\begin{equation}
\label{IIIA}
A_{jl}(0;z)= \int \dd w \int_{-\infty}^{w}\dd u\; \exp \{ [S(u;z)-S(w;z)]/\epsilon \}\, w^{j}\, u^{l} \Phi(u) \; .
\end{equation}
Analogous formula holds for $A^{\text{\tiny (in)}}_{jl}(0;z)$ 
with the Maxwell distribution $\Phi(u)$ replaced by the initial 
condition $G(u;0)=G_{\text{\tiny in}}(u)$.
Once system (\ref{III8}) has been solved, the insertion of the resulting expressions for $\tilde{N}_{1}(z)$ and $\tilde{N}_{2}(z)$ into 
formula (\ref{III6}) yields eventually an explicit solution of the kinetic equation for the velocity distribution
\begin{multline}
\label{III9}
\tilde{G}(w;z) = \frac{\Phi(w)}{z} + \frac{1}{\epsilon}\int_{-\infty}^w \dd u \, \exp \left\{ [S(u;z)-S(w;z)]/{\epsilon}\right\} \\
\times\left\{ G_{\text{\tiny in}}(u) + [A_{\epsilon}(z)\, u - B_{\epsilon}(z) ]\Phi(u) \right\} \; .
\end{multline}
With the shorthand notations $A_{jl}(z) =
 A_{jl}(0;z)$ and $A^{\text{\tiny (in)}}_{jl}(z)=A^{\text{\tiny (in)}}_{jl}(0;z)$, the formulae for coefficients 
$A_{\epsilon}(z)$ and $ B_{\epsilon}(z)$ read
\begin{equation}
\label{Aepsilon}
A_{\epsilon}(z)= \frac{1}{\Delta(z)} \left[\frac{\epsilon^2}{z} A_{00}(z)-2A_{00}(z)A_{10}^{\text{\tiny (in)}}(z) 
+2A_{10}(z)A_{00}^{\text{\tiny (in)}}(z)\right]
\end{equation}
and 
\begin{equation}
\label{Bepsilon}
B_{\epsilon}(z) = \frac{1}{\Delta(z)}
\left[\frac{\epsilon^2}{z} A_{01}(z)+ \epsilon A_{00}^{\text{\tiny (in)}}(z) 
+  2A_{11}(z)A_{00}^{\text{\tiny (in)}}(z) 
-2A_{01}(z) A_{10}^{\text{\tiny (in)}}(z)\right] \; ,
\end{equation}
where $\Delta(z)$, in accordance with the definition given in (\ref{C9}), 
is 
\begin{equation}
\label{III22}
\Delta (z)= \epsilon A_{00}(z) + 2\,\left( A_{00}(z)A_{11}(z)-A_{10}(z)A_{01}(z)\right) \, .
\end{equation}

\subsection{Stationary solution}

At large times, $\tau \to \infty$, we expect the velocity distribution to 
reach some stationary state  $G_{\text{\tiny st}}(w) = G(w;\infty)$. This 
can be easily checked by investigating the behaviour of $\tilde{G}(w;z)$ in the neighbourhood of $z=0$ at fixed velocity $w$. 

\bigskip

All integrals over $u$ in formula (\ref{III9}) do converge for any complex 
value of $z$. Moreover, all their derivatives with respect to $z$ are also well 
defined, as shown by differentiation under the integral sign. Thus, such integrals are entire 
functions of $z$. The sole quantities in expression (\ref{III9}) 
which become singular at $z=0$ 
are the coefficients $A_{\epsilon}(z)$ and $B_{\epsilon}(z)$, and obviously 
the term $\Phi(w)/z$. 
In fact, both $A_{\epsilon}(z)$ and $B_{\epsilon}(z)$ exhibit simple poles at 
$z=0$. Hence, the stationary solution of the kinetic 
equation (\ref{III2}) does emerge when $\tau \to \infty$, and it 
is given by the residue of the simple pole of $\tilde{G}(w;z)$ 
at $z=0$, namely  
\begin{equation}
\label{III10}
G_{\text{\tiny st}}(w) = \Phi(w) + \frac{\epsilon  }{\Delta(0)}
\int_{-\infty}^w \dd u \exp \left[\frac{S(u;0)-S(w;0)}{\epsilon}\right][ A_{00}(0)\, u\, - A_{01}(0) ]\Phi(u) \, . 
\end{equation}
In that expression, $A_{ij}(0)$ and $\Delta(0)$ are the non-zero values at $z=0$
of the analytic functions $A_{ij}(z)=A_{ij}(0;z)$ and $\Delta(z)=\Delta(0;z)$.
Formula (\ref{III10}) does not depend on initial condition 
$G_{\text{\tiny in}}$. All initial conditions evolve towards 
the same unique stationary distribution (\ref{III10}). It can be 
checked that the direct resolution of the static version of 
kinetic equation (\ref{III2}) obtained by setting 
$\partial G/\partial \tau = 0$ does provide formula (\ref{III10}). 

\bigskip

\begin{figure}
\includegraphics[width=0.9\textwidth]{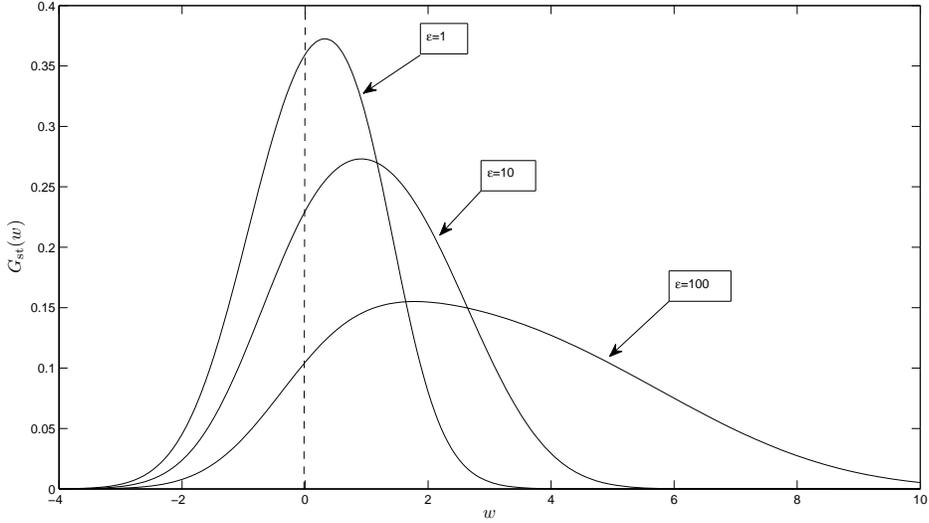}
\caption{\label{AP09b} Stationary velocity distribution 
$G_{\text{\tiny st}}(w)$ for three values of $\epsilon$.}
\end{figure}

Since the external field accelerates the particle, the stationary solution 
is asymmetric with respect to the reflection $w \rightarrow -w$, and positive 
velocities are favoured. This leads to a finite current
\begin{equation} 
\langle w \rangle_{\text{\tiny st}} = 
\int \dd w\, w \, G_{\text{\tiny st}}(w) = 
\frac{\epsilon }{\Delta(0)} [A_{00}(0)A_{11}(0)-A_{01}(0)A_{10}(0) ] \; .
\label{III11}
\end{equation}  
The asymptotic expansion at large velocities
of $G_{\text{\tiny st}}(w)$, inferred
from formula (\ref{III10}), reads
\begin{equation} 
G_{\text{\tiny st}}(w) = \frac{1}{\sqrt{2\pi}}e^{-w^{2}/2} 
\left[ 1 +  \frac{\epsilon^{2} A_{00}(0)}{\Delta(0)w} + O(\frac{1}{w^2}) 
\right]
\,\,\,\,\, \text{when} \,\,\,\,\, |w| \to \infty \, .
\label{III12}
\end{equation}  
Therefore, the external field does not influence the leading large-velocity
behaviour of $G_{\text{\tiny st}}(w)$, which is identical to 
that of the thermal bath. Its effects only 
arise in the first correction to the leading behaviour which is 
smaller by a factor of order $1/w$. 
The stationary distribution is drawn in Fig.~\ref{AP09b} for 
several increasing field strengths, $\epsilon = 1$, $\epsilon = 10$ and 
$\epsilon = 100$. 

\bigskip

Let us study now the limit $\epsilon \to 0$ which corresponds 
to a weak external field. The main contributions to the
integrals over $u$ in (\ref{III9}) arise from the region close to $w$. That observation 
motivates the use of a new integration variable $y=(w-u)/\epsilon$.
The Taylor expansions of the resulting integrands in powers of $\epsilon$ generate then
entire series in $\epsilon$, the first terms of which read 
\begin{equation}
\label{III13}
\int_{-\infty}^w \dd u \, u\, \Phi(u)
\exp \left[\frac{S(u;0)-S(w;0)}{\epsilon}\right] =
\epsilon \, \frac{w \Phi(w)}{1+w^2} + O(\epsilon^2)
\end{equation}
and 
\begin{equation}
\label{III14}
\int_{-\infty}^w \dd u \, \Phi(u)
\exp \left[\frac{S(u,0)-S(w,0)}{\epsilon}\right] =
\epsilon \, \frac{\Phi(w)}{1+w^2} + O(\epsilon^2) \, .
\end{equation} 
Consequently, also functions 
$A_{ij}(0)$ and $\Delta(0)$ can be represented by power series 
in $\epsilon$ as they are obtained by calculating appropriate moments of expansions (\ref{III13}) and (\ref{III14}) over the velocity space.
The corresponding small-$\epsilon$
expansion of the stationary velocity distribution reads
\begin{equation}
\label{III15}
G_{\text{\tiny st}}(w) = \Phi(w) + \epsilon \left[ \frac{b\, w }{1+w^2} \right]\Phi(w)
+  O(\epsilon^2) \, ,
\end{equation} 
where 
\[ b = \left[ 1+2 \int \dd w\, \frac{w^2}{1+w^2}\Phi(w)  \right]^{-1} \; . \] 
Of course, at $\epsilon = 0$,  $G_{\text{\tiny st}}(w)$ reduces to 
the Maxwell distribution. The first correction is 
of order $\epsilon$, as expected from linear response theory. 
The corresponding current (\ref{III10}) reduces to 
\begin{equation} 
\langle w \rangle_{\text{\tiny st}} = \sigma \epsilon 
+  O(\epsilon^2) \, ,
\label{III16}
\end{equation}  
where the conductivity $\sigma$ is given by
\begin{equation} 
\sigma = \frac{1}{2}( 1 - b )
\label{III17}
\end{equation}   
It will be shown in the sequel that $\sigma = D_0= D(\epsilon = 0)$, 
where $D(\epsilon) $ is the diffusion coefficient given by the Green-Kubo formula.

\bigskip

\begin{figure}
\includegraphics[width=0.9\textwidth]{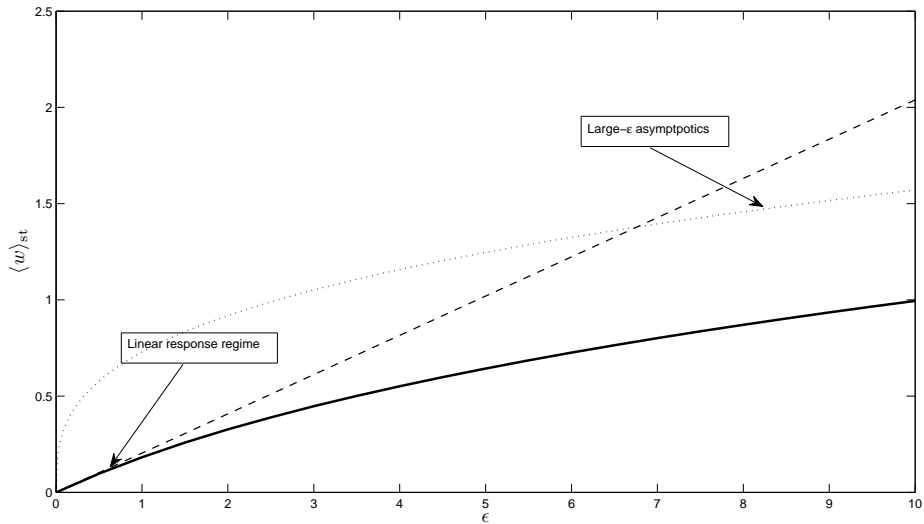}
\caption{\label{AP09c} Average current 
$\langle w \rangle_{\text{st}}$ as a function of $\epsilon$. 
The dashed line represents the linear Kubo term in the small-$\epsilon$ 
expansion (\ref{III16}) with 
conductivity $\sigma \simeq 0.2039$. The dotted line describes 
asymptotic formula (\ref{III20}) with 
$3^{1/3} \Gamma(2/3)/\Gamma(1/3) \simeq 0.7290$ valid in the limit 
$\epsilon \to \infty$.}
\end{figure}

Consider now the strong field limit $\epsilon \to \infty$. The 
corresponding behaviours of $A_{ij}(0)$ and $\Delta(0)$ are derived from the 
integral representations obtained in Appendix~\ref{C}. We then 
find at fixed $w$
\begin{equation}
\label{III18}
G_{\text{\tiny st}}(w) = \frac{\epsilon^{-1/3}}{\int_0^\infty \dd y \exp (-y^3/3)}
\int_{-\infty}^w \dd u \, \Phi(u)
\exp \left[\frac{S(u,0)-S(w,0)}{\epsilon}\right] + O(\epsilon^{-2/3}) 
\end{equation}
For $w$ of order 1,  the dominant term in the large-$\epsilon$ expansion of the integral in (\ref{III18}) reduces to  
\[  \int_{-\infty}^{w}\dd u \, \Phi(u)=\frac{1}{2}\left(1+\text{Erf}\left(\frac{w}{\sqrt{2}}\right)\right) \] 
and thus varies from $0$ to $1$ around the origin $w=0$.
For larger values of the velocity, $w \sim \epsilon^{1/3}$, 
that integral behaves as 
$\exp (-w^3/(3\epsilon)$. The next term
in the expansion (\ref{III18}) remains of order $\epsilon^{-2/3}$. 
Thus, when $\epsilon \to \infty$ at fixed $\epsilon^{-1/3} w$ the stationary solution is given by
\begin{equation}
\label{III19}
G_{\text{\tiny st}}(w) \sim \theta(w)\, \frac{\epsilon^{-1/3}}{\int_0^\infty \dd y \exp (-y^3/3)}
\, \exp \left[-(\epsilon^{-1/3}w)^3/3 \right]  \, , 
\end{equation}
 where $\theta$ is the Heaviside step function. The whole distribution is shifted toward 
high velocities $w \sim \epsilon^{1/3}$, so that the
resulting current (\ref{III11}) is of the same order of magnitude, 
\textsl{i.e.} 
\begin{equation} 
\langle w \rangle_{\text{\tiny st}} \sim \frac{3^{1/3} \Gamma(2/3)}{\Gamma(1/3)} \; \epsilon^{1/3} \,\,\,\,  
\text{when} \,\,\,\, \epsilon \to \infty \, ,
\label{III20}
\end{equation}  
where $\Gamma$ is the Euler Gamma function.
That behavior can be recovered within the following simple interpretation.
At strong fields, the average velocity of the particle becomes large compared 
to the thermal velocity of scatterers. Since 
at each collision the particle exchanges its velocity with a thermalized
scatterer, the variation of particle velocity between two successive collisions is 
of the order of $\langle v \rangle_{\text{\tiny st}}$.
On the other hand, in the stationary state the same velocity variation is due to the acceleration $a$ 
coming from the external field, so it is of the order 
$a \tau_{\text{\tiny coll}}$ where $\tau_{\text{\tiny coll}}$ 
is the mean time between two successive collisions. 
This time can be reasonably 
estimated as the inverse collision frequency 
for a relative velocity $|v-c|$ of order $\langle v \rangle_{\text{\tiny st}}$. 
The consistency of those estimations requires the relation
\begin{equation} 
\langle v \rangle_{\text{\tiny st}} \sim a \; \frac{v_{\text{\tiny int}}}
{\rho \, \langle v \rangle_{\text{\tiny st}}^2} \,
\label{III21}
\end{equation} 
which indeed implies the $\epsilon^{1/3}$-behaviour 
(\ref{III20}) of the average velocity in dimensionless units.
Contrary to the Maxwell case where the current 
remains linear in the applied field, here the current deviates 
from its linear-response form when the field increases : it grows 
more slowly because collisions are more efficient in dissipating 
the energy input of the field. In Fig.~\ref{AP09c}, we plot 
$\langle w \rangle_{\text{\tiny st}}$ as a function of $\epsilon$.

\subsection{Relaxation towards the stationary solution}

Let us study now the relaxation of the velocity distribution 
$G(w;\tau)$ towards the stationary solution $G_{\text{\tiny st}}(w)$. 
The decay of $[ G(w,\tau) - G_{\text{\tiny st}}(w) ]$ when $\tau \to \infty$
is controlled by the singularities of $\tilde{G}(w;z)$ 
in the complex plane, different from the pole at $z=0$. 
As already mentioned, all integrals in expression (\ref{III9}) 
are entire functions of $z$, so the singularities at $z \neq 0$ 
arise only in the
coefficients $A_{\epsilon}(z)$ and $B_{\epsilon}(z)$. Thus, the first important 
conclusion is that the relaxation is uniform for the whole velocity spectrum.

\bigskip

According to expressions (\ref{Aepsilon}) and (\ref{Bepsilon}) 
defining $A_{\epsilon}(z)$ and $B_{\epsilon}(z)$ respectively, 
the singularities of those coefficients at points $z\neq 0$, 
correspond to zeros of the function $\Delta(z)$ given by expression
(\ref{III22}). Since the analytic functions $A_{ij}(z)$ and 
$\Delta (z)$ do not depend on initial condition 
$G_{\text{\tiny in}}$. the relaxation is an intrinsic dynamical process, as expected. 

\bigskip

After some algebra detailed in Appendix~\ref{C}, 
we find that $\Delta (z)$ reduces to
the Laplace transform 
\begin{equation}
\label{III23}
\Delta (z)=\epsilon^{2} \, \int_0^{\infty} \dd y f_{\epsilon}(y)
\exp (-zy)  
\end{equation}
of the real, positive, and monotonously decreasing function
\begin{equation}
\label{III24}
\epsilon^{2} f_{\epsilon}(y)=  \frac{\epsilon^{2} (1+3y)}{(1+y)(1+2y)^{1/2}}
\exp \left( -y -\epsilon^2 \frac{y^3(2+y)}{6(1+2y)} \right)   \, .
\end{equation}
Owing to the fast 
decay of $f_{\epsilon}(y)$ the integral (\ref{III23}) converges for any $z$, so 
$\Delta (z)$ is an entire function of $z$. Also, the monotonic decay of $f_{\epsilon}(y)$ 
and its positivity imply some general properties for the roots of $\Delta (z)$.
First of all, $\Delta (z)$ cannot vanish for $\Re (z) \geq 0$. 
Moreover, as $\Delta (z)$ is strictly positive 
for $z$ real, the zeros of $\Delta (z)$ appear in complex conjugate pairs, 
while they are isolated with strictly negative real parts 
and nonvanishing imaginary parts. Consequently, the long-time relaxation of the velocity distribution is governed by
the pair of zeros which is closest to the imaginary axis. Noting them as
 $z^{\pm}=-\lambda \pm i \omega$ with  $\omega \neq 0$ and $0 < \lambda$, we conclude that
$G(w;\tau)$ relaxes towards $G_{\text{\tiny st}}(w)$
via exponentially damped oscillations
\begin{equation}
\label{III25}
G(w;\tau) -  G_{\text{\tiny st}}(w) \sim C(w) \cos [\omega \tau + \eta(w)] 
\exp (-\lambda \tau), \,\,\,\,  
\text{when} \,\,\,\, \tau \to \infty \, 
\end{equation}
where $C(w)$ and $\eta(w)$ are an amplitude and a
phase respectively. It should be noticed that both functions 
$C(w)$ and $\eta(w)$ depend on initial conditions.

\bigskip

At a given value of $\epsilon$, the zeros $z^\pm$ are found
by solving numerically the equation $\Delta (z^\pm)=0$. 
In the weak- or strong-field limits, we can derive 
asymptotic formulae for such zeros as follows. 
First, as indicated by numerically computing $z^\pm$ for small values of $\epsilon$,
$z^\pm$ collapse to $z_0=-1$ when $\epsilon \to 0$. The corresponding asymptotical  
behaviour can be derived by noting that, for $z$ close to 
$z_0$, the leading contributions to $\Delta(z)$ in integral 
(\ref{III23}) arise from large values of $y$. Then, we set 
$y=\xi/\epsilon^{2/3}$ and $z=-1 +s\; \epsilon^{2/3}$, which 
provide
\begin{equation}
\label{III26bis}
\Delta (-1 +s\; \epsilon^{2/3}) \sim \frac{3\;\epsilon^{5/3}}{\sqrt{2}} \, 
\int_0^\infty \dd \xi \; \xi^{-1/2} \exp(-s\;\xi-\xi^3/12)  
\end{equation}
when $\epsilon \to 0$ at fixed $s$. By numerical methods, we find the 
pair of complex conjugated zeros $s_0^\pm $ of integral
\[ \int_0^\infty \dd \xi \; \xi^{-1/2} \exp(-s\;\xi-\xi^3/12) \]
which are the closest to the imaginary axis. Therefore, when $\epsilon \to 0$, 
damping factor $\lambda(\epsilon)$ goes to $1$ according to 
\begin{equation}
\label{III26ter}
\lambda(\epsilon)=1-\Re(s_0^\pm)\;\epsilon^{2/3} +o(\epsilon^{2/3})
\end{equation}
with $\Re(s_0^\pm) \simeq -1.169$, while frequency $\omega(\epsilon)$ 
vanishes as $\Im(s_0^+)\;\epsilon^{2/3}$ with $ \Im(s_0^+) \simeq 2.026$.
Notice that for fixed $z$, not located on the real half-axis $]-\infty,-1]$,
$\Delta(z)$ behaves as 
\begin{equation}
\label{III26}
\Delta (z) \sim \epsilon^{2} \, \Delta_0(z) 
\end{equation}
when $\epsilon \to 0$, with
\begin{multline}
\label{III27}
\Delta_0(z) = \sqrt{\frac{\pi}{2(z+1)}} \,  e^{(z+1)/2}
\left[ 1 - \text{Erf}\left(\sqrt{(z+1)/2}\right)\right] \\
\times \left[3-\sqrt{2\pi(z+1)} e^{(z+1)/2} \left(1 - \text{Erf}\left(\sqrt{(z+1)/2}\right)\right)\right] \, .
\end{multline}
Here, $\sqrt{(z+1)/2}$ is defined as the usual real positive 
square root $\sqrt{(x+1)/2}$
for real $z=x$ belonging to the half axis $x > -1$,  
while the complementary half-axis
$z=x \leq -1$ is a cut-line ending at the branching point $z=-1$. That point 
is the singular point of $1/\Delta_0(z)$ closest to the imaginary axis, as 
strongly suggested by a numerical search of the zeros of $\Delta_0(z)$. Therefore, 
both $\lambda(\epsilon)$ and $\omega(\epsilon)$ are continuous functions of 
$\epsilon$ at $\epsilon=0$ with $\lambda(0)=1$ and $\omega(0)=0$. At $\epsilon=0$, the exponentially damped oscillating decay (\ref{III25}) becomes 
an exponentially damped monotonic decay multiplied by power-law $t^{-3/2}$. 
That power-law arises from the presence of a singular term of order 
$\sqrt{(z+1)/2}$ in the expansion of $\tilde{G}(w;z)$ around the branching 
point $z=-1$.

\bigskip

\begin{figure}
\includegraphics[width=0.9\textwidth]{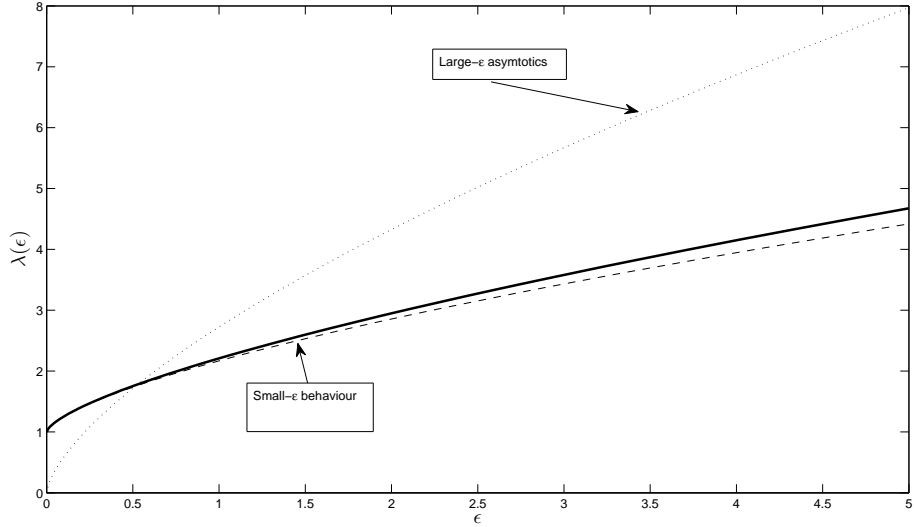}
\caption{\label{AP09e} Damping factor
$\lambda(\epsilon )$ as a function of $\epsilon$. The dashed and 
dotted lines represent the asymptotical behaviours 
(\ref{III26ter}) and (\ref{III28bis}) at small and large $\epsilon$ respectively.}

\end{figure}

When $\epsilon \to \infty$, the zeros of $\Delta (z)$ 
are obtained by simultaneously changing $y$ to 
$\xi/\epsilon^{2/3}$ in the integral (\ref{III23}) and by rescaling 
$z$ as $\epsilon^{2/3}s$. This provides 
\begin{equation}
\label{III28}
\Delta (\epsilon^{2/3}s) \sim \epsilon^{4/3} \, \Delta_\infty (s) 
\,\,\,\,\text{when}\,\,\,\, \epsilon \to \infty\,\,\,\,\text{at fixed}\,\,\,\, s\, ,
\end{equation}
with 
\begin{equation}
\label{III29}
\Delta_\infty (s)= \int_0^{\infty} \dd \xi \exp \left(-s\, \xi -\xi^3/3 \right)\, .
\end{equation}
Therefore, when $\epsilon \to \infty$, $z^{\pm}$ behave as 
$z^{\pm} \sim \epsilon^{2/3}s_\infty^\pm$, where $s_\infty^\pm$ are the 
zeros of $\Delta_\infty (s)$ closest to the imaginary axis. The corresponding 
large-$\epsilon$ asymptotical behaviour of the damping factor
$\lambda(\epsilon)$ is 
\begin{equation}
\label{III28bis}
\lambda(\epsilon)=-\Re(s_\infty^\pm)\;\epsilon^{2/3} + o(\epsilon^{2/3})
\end{equation}
with $\Re(s_\infty^\pm) \simeq -2.726$, while frequency $\omega(\epsilon)$ 
diverges as $\Im(s_\infty^+)\;\epsilon^{2/3}$ with $ \Im(s_\infty^+) \simeq 6.260$.
Notice that the relaxation time $\lambda^{-1}(\epsilon)$ goes to zero as $\epsilon^{-2/3}$,
like the average time between collisions 
$\tau_{\text{\tiny coll}} \sim \langle v \rangle_{\text{\tiny st}}/a$ 
used in our simple heuristic derivation of the $\epsilon$-dependence 
of the stationary current in the strong field limit.
In Fig.~\ref{AP09e}, we draw the damping factor $\lambda(\epsilon)$ as a function of $\epsilon$.

\subsection{Relaxation of density in position space}

In Appendix~\ref{A} we derive an explicit formula for the zeroth moment $\tilde{M}_{0}(k;z)$ of the distribution $\tilde{F}(k,w;z)$ which contains all information on the evolution of the spatial density of the propagating particle. The formula (\ref{C8}) clearly reveals the presence of a hydrodynamic pole in $\tilde{M}_{0}(k;z)$, namely the root of equation
\begin{equation}
\label{D1}
z + (k^2 + i\epsilon k)\; U(k;z) = 0
\end{equation}
where
\begin{equation}
\label{U}
U(k;z) = \frac{A_{11}(k;z)A_{00}(k;z)-A_{10}(k;z)A_{01}(k;z)}{\epsilon A_{00}(k;z)+ 2[A_{11}(k;z)A_{00}(k;z)-A_{10}(k;z)A_{01}(k;z)]} \; .
\end{equation}

\bigskip

If we consider the small-$k$ limit and if we assume the asymptotic form 
\begin{equation}
\label{D2}
z_{\rm{hy}}(k) = -ic k - D(\epsilon)\, k^2 + 0(k^2) 
\end{equation}
for the hydrodynamic root,
we find immediately from equation (\ref{D1}) the formula 
\begin{equation}
\label{D3}
c = \epsilon\; U(0;0) \; . 
\end{equation}
This shows that the mode propagates with the average stationary velocity $ \langle w \rangle_{\text{\tiny st}} = 
\epsilon \, U(0;0) $ derived in expression (\ref{III11}).

\bigskip

In order to infer the formula for the diffusion coefficient $D(\epsilon)$, 
it is necessary to calculate the term linear in variable $k$ 
in the expansion of function $U(k;z)$ at $z=-ick$. Indeed, equation 
(\ref{D1}) implies the equality
\begin{equation}
\label{D4}
D(\epsilon  ) = U(0;0) + i\epsilon \, \frac{\dd}{\dd k}U(k;-ick)|_{k=0}  \; . 
\end{equation}
Taking into account the structure (\ref{D4}) of $U(k;z)$ we find the formula
\begin{equation}
\label{D5}
D(\epsilon ) = \frac{\langle w \rangle_{\text{\tiny st}}}{\epsilon} + \frac{ A_{00}[{A}^{\prime}_{11}A_{00}- {A}^{\prime}_{01}A_{10}]+ A_{01}[ {A}^{\prime}_{00}A_{10} -{A}^{\prime}_{10}A_{00}]}{\Delta^{2}} 
\end{equation}
where all $A_{jl}$ and $\Delta$ are taken at $k=z=0$, and where
\begin{equation}
\label{D6}
{A}^{\prime}_{jl} = i\epsilon \frac{\dd}{\dd k}A_{jl}(k;-ick)|_{k=0} \; .
\end{equation}
A particularly useful representation of the derivative appearing in 
expression (\ref{D6}) can be deduced from  formulae (\ref{S}) and (\ref{C5}) defining functions $A_{jl}(k;z)$. An integration by parts yields
\begin{equation}
\label{D7}
{A}^{\prime}_{jl}=\int \dd w \int^{w}_{-\infty} \dd u\, (u-c) \int^{u}_{-\infty} \dd v \,
w^j\, v^l\exp \{ [S(0,v;0)-S(0,w;0)]/\epsilon \}\Phi(v) \; .
\end{equation} 
It is quite remarkable that equation (\ref{D7}) allows us 
to establish a relation between the quantities ${A}^{\prime}_{jl} $ 
and the stationary velocity distribution $G_{\text{\tiny st}}(w)$. 
Indeed, using equation (\ref{III10}), we  readily obtain the equalities
\begin{multline}
\int \dd w \int_{-\infty}^{w} \dd u \exp \{ [S(0,v;)-S(0,w;0)]/\epsilon \}(u-c)\; G_{\text{\tiny st}}(u) \\
=  A_{01}-c A_{00} + \frac{1}{\Delta} [A_{00}{A}^{\prime}_{01} - A_{01}{A}^{\prime}_{00} ] \equiv J_{01}
\label{D8a}  
\end{multline}
and
\begin{multline}
\int \dd w \int_{-\infty}^{w} \dd u \exp \{ [S(0,v;)-S(0,w;0)]/
\epsilon \}\,w\, (u-c)\; G_{\text{\tiny st}}(u) \\
=  A_{11}-c A_{10} + \frac{1}{\Delta} [A_{00}{A}^{\prime}_{11} - A_{01}{A}^{\prime}_{10} ]\equiv J_{11} \; .
\label{D9}   
\end{multline} 
Then, we find that the linear combination $(A_{00}J_{11}-A_{10}J_{01})$ 
of integrals $J_{11}$ and $J_{01}$ reduces to
\begin{equation}
\label{D10}
A_{11}A_{00}-A_{10}A_{01} + \frac{1}{\Delta} \left\{ A_{00}[A_{00}{A}^{\prime}_{11} - A_{01}{A}^{\prime}_{10}]-
A_{10}[A_{00}{A}^{\prime}_{01} - A_{01}{A}^{\prime}_{00} ]   \right\} \; .
\end{equation}
The comparison of that expression with equation (\ref{D5}) 
leads to the compact final result
\begin{equation}
\label{D11}
D(\epsilon ) = \frac{A_{00}J_{11}-A_{10}J_{01}}{\Delta} \; .
\end{equation}
The above formula involves, \textsl{via} coefficients $J_{11}$ 
and $J_{01}$, averages over the stationary velocity distribution. In fact, we show in Appendix~\ref{B} that expression (\ref{D11}) follows 
by extending, to the present out-of-equilibrium stationary 
state, the familiar Green-Kubo relation between the diffusion 
coefficient and the velocity fluctuations. That important fact is one of 
the main observations of the present study.

\bigskip

\begin{figure}
\includegraphics[width=0.9\textwidth]{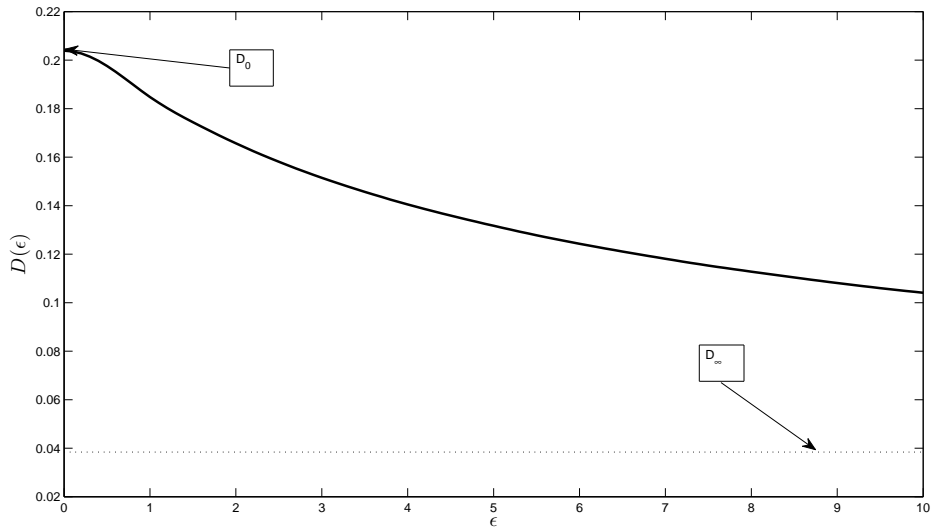}
\caption{\label{AP09d} Diffusion coefficient 
$D(\epsilon )$ as a function of $\epsilon$. The dotted line 
represents the constant asymptotic value 
$D_{\infty}$.}
\end{figure}

When $\epsilon \to 0$, the behaviour of $D(\epsilon)$ is easily infered 
by inserting the small-$\epsilon$ expansion (\ref{III15}) of 
the stationary velocity distribution $G_{\text{\tiny st}}(w)$ into
formula (\ref{D11}). We find that $D(\epsilon)$ goes to conductivity 
$\sigma$ (\ref{III17}) as quoted above, with a negative 
$\epsilon^2$-correction.
When $\epsilon \to \infty$, we can use the large-$\epsilon$ form 
(\ref{III19}) of $G_{\text{\tiny st}}(w)$ for evaluating coefficients $J_{11}$ and $ J_{01}$.
Using also the corresponding behaviours of coefficients $A_{00}$ and $A_{10}$, 
we eventually obtain that $D(\epsilon)$ goes to the finite 
value 
\begin{equation}
\label{D12}
D_\infty = \frac{\Gamma^3(1/3)-9\Gamma(1/3)\Gamma(2/3)+6\Gamma^3(2/3)}{2 \Gamma^3(1/3)} \simeq 0.0384 \; .
\end{equation}
The external field dependence of the diffusion coefficient $D(\epsilon)$
is shown in Fig.~\ref{AP09d}. 

\bigskip

The expansion (\ref{D2}) of $z_{\rm{hy}}(k)$ can be pursued beyond 
the $k^2$-diffusion term, by expanding function $U(k;z)$ in double 
entire series with respect to $z$ and $k$. According to the integral 
expression of functions  $A_{jl}(k;z)$ derived in Appendix~\ref{C}, all 
coefficients of those double series are finite. This implies that 
the hydrodynamic root $z_{\rm{hy}}(k)$ of equation (\ref{D1}) can be 
formally represented by an entire series in $k$, namely
\[ z_{\rm{hy}}(k)=\sum_{n=1}^\infty \alpha_n k^n \; ,\]
with $\alpha_1=-ic$ and $\alpha_2=-D(\epsilon)$. Coefficient $\alpha_n$ 
($n \geq 3$) can be straightforwardly computed once lowest-order 
coefficients $\alpha_p$ with $1 \leq p \leq n-1$ have been determined. 
As shown by that calculation, all coefficients are obviously finite. 
Therefore, and similarly to what happens in the Maxwell case, only  
positive integer powers of $k$ appear in the small-$k$ expansion 
of $z_{\rm{hy}}(k)$. Now, we are not able to determine the radius of 
convergence of that expansion, so we cannot conclude about the analyticity of 
function $z_{\rm{hy}}(k)$. However, we notice that, contrarily to the 
Maxwell case, the integrals defining $A_{jl}(k;z)$ remain well-defined 
for any complex value of $k$, as soon as $\epsilon \neq 0$ 
(see Appendic~\ref{C}). This suggests 
that $z_{\rm{hy}}(k)$ might be an analytic function of $k$ at $k=0$, 
except for $\epsilon = 0$, in which case $k=0$ should be a singular point.

\section{Concluding comments}

The idea of this work was to perform a detailed study of the 
approach to an out-of-equilibrium stationary state, 
by considering systems for which analytic solutions 
can be derived. To this end we solved, within Boltzmann's kinetic theory, 
the one-dimensional initial value problem for the distribution of a particle accelerated by a constant external field and suffering elastic collisions with thermalized bath particles. Our exact results for the Maxwell model and 
for the very hard particle model
support the general picture mentioned in the Introduction:
\begin{itemize}
\item a uniform exponentially fast relaxation of the velocity distribution 
\item diffusive spreading in space in the reference system moving with stationary flow
\item equality between the diffusion coefficient appearing in the hydrodynamic mode and the
      one given by the generalized Green-Kubo formula 
\end{itemize}

\bigskip

Although both models display the same phenomena listed above, the  
variations of the respective quantities of interest with respect 
to $\epsilon$ are different. First we notice that, as far as deformations 
of the equilibrium Maxwell distribution are concerned, the 
external field is much less efficient for very hard particles. This is 
well illustrated by comparing figures \ref{AP09a} and \ref{AP09b} : 
for the Maxwell system, a significative deformation of $\Phi$ is 
found for $\epsilon=5$, while for the very-hard particle model a similar 
deformation is observed for $\epsilon=100$. 
This can be easily interpreted as follows.
The collision frequency for very hard particles 
becomes much larger than its Maxwell gas counterpart when the 
external field increases, so it costs more energy to maintain 
a stationary distribution far from the equilibrium one. 
That mechanism also explains various related 
observations. For instance, the large-velocity behaviour of 
$G_{\text{\tiny st}}(w)$ is identical to the equilibrium Gaussian 
for very hard particles, while it takes an exponential form in the Maxwell
gas. Also, the average current $\langle w \rangle_{\text{\tiny st}}$ 
increases more slowly when $\epsilon \to \infty$ for very hard particles, 
and the corresponding relaxation time $\lambda^{-1}(\epsilon)$ 
vanishes instead of remaining constant for the Maxwell gas.

\bigskip

Among the above phenomena, the emergence of a symmetric diffusion process 
in the moving reference frame is quite remarkable. In such a frame, there 
is some kind of cancellation between the action of the external field and 
the effects of collisions induced by the counterflow of bath particles 
with velocity  $u_{\text{\tiny bath}}^{\ast}=-
\langle v \rangle_{\text{\tiny st}}$. The corresponding diffusion coefficient
$D(\epsilon)$ increases whith $\epsilon$ 
for the Maxwell gas (case $\gamma=0$), 
while it decreases and 
saturates to a finite value for very hard particles 
(case $\gamma=2$). Therefore, beyond the 
previous cancellation, it seems that the large number of collisions 
for $\gamma=2$ shrink equilibrium fluctuations. On the contrary, for 
$\gamma=0$, since $D(\epsilon)$ diverges when 
$\epsilon \to \infty$, the residual effect of collisions 
in the reference frame seems to vanish and particles tend to behave 
as if they were free. 

\bigskip

We expect that the same qualitative picture should be valid in the hard rod case which corresponds to the intermediate value  $\gamma = 1$ of the exponent 
$\gamma$ in equation (\ref{I1}). The quantitative behaviours 
should interpolate between those described for $\gamma=0$ and 
$\gamma=2$. For instance, the stationary distribution 
$G_{\text{\tiny st}}(w)$ computed in Ref.~\cite{GP86} displays 
a large-velocity asymptotic behaviour which is indeed intermediary between 
those derived here for $\gamma=0$ and $\gamma=2$. Also, the 
average current $<v>_{\text{\tiny st}}$ is of order $\epsilon^{1/2}$ for 
$\epsilon$ large, which lies between the $\epsilon$- and 
$\epsilon^{1/3}$-behaviours found for $\gamma=0$ and $\gamma=2$
respectively. Notice that the $\epsilon^{1/3}$-behaviour for $\gamma=2$
can be retrieved within a selfconsistent argument, which uses 
in an essential way the existence of the velocity scale related to the particle-particle interaction. Whereas the thermal velocity scale becomes irrelevant when $\epsilon\to\infty$, the interaction scale remains important. 
In the case of hard rods such an interaction scale does not show 
in the kinetic equation, and the unique combination of parameters 
having the dimension of velocity is $\sqrt{a/\rho}$, 
which does provide a different strong field behaviour of 
$<v>_{\text{\tiny st}}$ with order $\epsilon^{1/2}$.

\appendix

\section{Solution of the kinetic equation for very hard particles}
\label{A}

Applying to equation (\ref{III1}) Fourier and  
Laplace transformations, we find
\begin{multline}
\epsilon\frac{\partial}{\partial w}\tilde{F}(k,w;z) + 
(z + 1 + ikw + w^{2}) \tilde{F}(k,w;z) \\
=\hat{F}_{\text{\tiny in}}(k,w) + 
[\tilde{M}_{2}(k;z)-2w\tilde{M}_{1}(k;z)+ w^{2}\tilde{M}_{0}(k;z)] \Phi(w)
\label{C1}
\end{multline}
where $\tilde{M}_{j}(k;z)$ is the double Fourier-Laplace transform of the 
$j^{th}$-moment $M_{j}(x;\tau)$ defined in expression (\ref{IIIM}), while $\hat{F}_{\text{\tiny in}}(k,w)$ 
is the spatial Fourier transform of the initial condition 
$F_{\text{\tiny in}}(x,w)= F(x,w;0)$. 
The first order equation (\ref{C1}) can be rewritten in an integral form with the use of function
\begin{equation}
\label{S}
S(k,w;z)= w(z+1)+ \frac{1}{3}w^3 + ik\frac{w^2}{2} \; ,
\end{equation}
namely
\begin{multline}
\epsilon \tilde{F}(k,w;z) = \int_{-\infty}^{w}\dd u \exp \{ [S(k,u;z)-S(k,w;z)]/\epsilon \}\; \{ \hat{F}_{\text{\tiny in}}(k,u)  \\
+ [\tilde{M}_{2}(k;z)-2u\tilde{M}_{1}(k;z)+ u^{2}\tilde{M}_{0}(k;z)] \Phi(u) \}
\; .
\label{C2} 
\end{multline}
Using then the relation
\[ \left[ \epsilon \frac{\partial}{\partial u}+\epsilon u -(z+1) -iku - u^2 \right] \exp \{ S(k,u;z)/\epsilon \}\Phi (u) = 0\]
to evaluate the term involving $u^2$ in the right hand side of 
equation (\ref{C2}), we eventually find the more convenient 
integral equation
\begin{multline}
\epsilon \tilde{F}(k,w;z) = \epsilon \tilde{M}_{0}(k;z)\Phi(w) + 
\int_{-\infty}^{w}\dd u \exp \{ [S(k,u;z)-S(k,w;z)]/\epsilon \} \\ 
\times \{ \hat{F}_{\text{\tiny in}}(k,u)  
+ [\tilde{M}_{2}(k;z)-2u\tilde{M}_{1}(k;z)+ (\epsilon u-iku -z-1)\tilde{M}_{0}(k;z)] \Phi(u) \} \; . 
\label{C3}
\end{multline}
Equation (\ref{C3}) has to be considered together with the continuity equation
\begin{equation}
\label{C4}
z \tilde{M}_{0}(k;z) + ik \tilde{M}_{1}(k;z) = \hat{M}_{0}(k;0) \, .
\end{equation}
In order to determine the unknown functions $\tilde{M}_{j}(k;z)$ ($j=0,1,2$), 
we complete equation (\ref{C4}) with the zeroth and the first moments of equation (\ref{C3}).  In the resulting system of linear equations, 
the integrals
\begin{equation}
\label{C5}
A_{jl}(k;z)= \int \dd w \int_{-\infty}^{w}\dd u\; w^j\, u^l\exp \{ [S(k,u;z)-S(k,w;z)]/\epsilon \}\Phi(u)
\end{equation}
and
\begin{equation}
\label{C6}
 A^{\text{\tiny (in)}}_{jl}(k;z)= \int \dd w \int_{-\infty}^{w}\dd u\; w^j\, u^l\exp \{ [S(k,u;z)-S(k,w;z)]/\epsilon \}\hat{F}_{\text{\tiny in}}(k,u)
\end{equation}
appear. That system reads
\begin{eqnarray}
\label{C7}
0 & = & A^{\text{\tiny (in)}}_{00} + [ \tilde{M}_{2} - (z+1)\tilde{M}_{0}]A_{00}+[(\epsilon -ik) \tilde{M}_{0}-2\tilde{M}_{1}]A_{01}\\
\epsilon \tilde{M}_{1} & = &  A^{\text{\tiny (in)}}_{10} + [ \tilde{M}_{2} - (z+1)\tilde{M}_{0}]A_{10}+[(\epsilon -ik) \tilde{M}_{0}-2\tilde{M}_{1}]A_{11}
\nonumber
\end{eqnarray}
The explicit solution for the zeroth moment reads
\begin{equation}
\label{C8}
\tilde{M}_{0}= \frac{\Delta \hat{M}_{0}(k;0)-ik[ A^{\text{\tiny (in)}}_{10}A_{00}-A_{10}A^{\text{\tiny (in)}}_{00} ]}{ z\Delta +
 (k^2 + i\epsilon k)(A_{11}A_{00}-A_{10}A_{01})}
\end{equation}
where
\begin{equation}
\label{C9}
\Delta = \epsilon A_{00} + 2(A_{11}A_{00}-A_{10}A_{01}) \; .
\end{equation}
 The formula for the first moment follows directly from the 
continuity equation (\ref{C4}), and then $\tilde{M}_{2}$ can be 
derived directly from (\ref{C7}). The insertion of the formulae for the 
first three moments into the relation (\ref{C3}) yields the complete solution 
for the distribution $\tilde{F}(k,w;z)$ for any initial condition.

\section{Evaluation of the diffusion coefficient via Green-Kubo theory}
\label{B}

\subsection{Velocity autocorrelation function of the Maxwell gas}

In order to evaluate the velocity autocorrelation function, 
we use the integral representation
\begin{equation}
\label{B0}
 \Gamma(\tau)= <[ w(\tau) - <w>_{\text{\tiny st}} ]
[w(0) - <w>_{\text{\tiny st}} ]>_{\text{\tiny st}}=
\int \dd w \; w \; G(w;\tau)  
\end{equation}
where $G(w;\tau)$ is the solution of kinetic equation (\ref{II4})  
corresponding to the initial condition
\begin{equation}
\label{B1}
G_{\text{\tiny in}}(w) = (w - <w>_{\text{\tiny st}})G_{\text{\tiny st}}(w) = (w - \epsilon)G_{\text{\tiny st}}(w)
\end{equation}
Here we find
\begin{equation}
\label{B2}
N_{0} = \int \dd w \; G(w;\tau)=\int \dd w \; G(w;0) = 0 \; ,
\end{equation}
so that equation (\ref{II4}) takes a particularly simple form
\begin{equation}
\label{B3}
G(w;\tau) = e^{-\tau} G_{\text{\tiny in}}(w-\epsilon \tau ) = e^{-\tau} (w - \epsilon) G_{\text{\tiny st}}(w-\epsilon \tau) \; . 
\end{equation}
Using then the explicit form (\ref{II5}) of $G_{\text{\tiny st}}$, we obtain
\begin{equation}
\label{B4}
\Gamma(\tau)=\int \dd w \; w e^{-\tau} (w - \epsilon)  \int_{0}^{\infty} \dd \eta e^{-\eta}\Phi(w-\epsilon\eta) = e^{-\tau}(1+\epsilon^2) \; , 
\end{equation}
which leads to the simple formula for the diffusion coefficient
\begin{equation}
\label{BD}
D(\epsilon)= \int_{0}^{\infty }\dd \tau \, \Gamma(\tau)= 1+\epsilon^{2} \; .
\end{equation}

\subsection{Velocity autocorrelation function of very hard particles}

Similarly to the Maxwell gas case, in order to determine the autocorrelation function $\Gamma(\tau)$, we first have to solve kinetic 
equation (\ref{III2}) satisfied by the velocity distribution 
with the initial condition 
$G(w;0) = G_{\text{\tiny in}}(w) = (w - <w>_{\text{\tiny st}})G_{\text{\tiny st}}(w) $, and afterwards we have to evaluate the first 
moment of that solution
\[  \int \dd w \, w \; G(w;\tau) = \Gamma (\tau) \; .\]
Since norm ${N}_{0}(\tau)$ vanishes, kinetic equation (\ref{III2}) 
becomes
\begin{equation}
\label{B5}
\left( \frac{\partial}{\partial \tau}+\epsilon\frac{\partial}{\partial w} \right)G(w;\tau) =
[ N_{2}(\tau)-2wN_{1}(\tau)]\Phi(w)-(w^{2}+1)G(w;\tau) \, .
\end{equation}
That equation can be rewritten in Laplace world as 
\begin{multline}
\epsilon \tilde{G}(w;z) = \int^{w}_{-\infty} \dd u \exp\left\{ [S(u;z)-S(w;z)]/\epsilon \right\} \\
\times \left\{ (u-<w>_{\text{\tiny st}}) G_{\text{\tiny st}}(u) +[ \tilde{N}_{2}(z) -2\tilde{N}_{1}(z)\, u ]\Phi (u)\right\} \; .
\label{B6}
\end{multline}
The zeroth and the first moments of equation (\ref{B6}) provide the system of equations
\begin{eqnarray}
\label{B7}
0 & = & J_{01} + \tilde{N}_{2}A_{00} -2\tilde{N}_{1}A_{01}\\
\epsilon \tilde{N}_{1} & = &  J_{11} + \tilde{N}_{2}A_{10}-2\tilde{N}_{1}A_{11}
\nonumber
\end{eqnarray}
where
\begin{equation}
\label{B8}
J_{01}=\int \dd w \int^{w}_{-\infty} \\d u \exp\left\{ [S(u;z)-S(w;z)]/\epsilon \right\}(u-<w>_{\text{\tiny st}}) 
G_{\text{\tiny st}}(u)
\end{equation}
and
\begin{equation}
J_{11}=\int \dd w \int^{w}_{-\infty} \\d u \exp\left\{ [S(u;z)-S(w;z)]/\epsilon \right\}w\, (u-<w>_{\text{\tiny st}}) 
G_{\text{\tiny st}}(u) \; . 
\end{equation}
The first moment $\tilde{N}_{1}(z)=\tilde{\Gamma}(z)$ is found to be
\begin{equation}
\label{B9}
\tilde{\Gamma} (z) = \frac{A_{00}(z)J_{11}(z)-A_{10}(z)J_{01}(z)}{\epsilon A_{00}(z) + 2 [A_{00}(z)A_{11}(z)-A_{10}(z)A_{01}(z)]} \; ,
\end{equation}
where the shorthand notation $A_{jl}(z)\equiv A_{jl}(0;z)$ has been used. The value of $\tilde{\Gamma}(z)$ at $z=0$ yields the diffusion coefficient $D(\epsilon)$. We find here the same formula as that derived from the analysis of the hydrodynamic pole (see expression (\ref{D11})).

\section{Useful integral expressions for functions arising in 
the case of very hard particles }
\label{C}

In the integral representation (\ref{C5}) of function $A_{jl}(k;z)$, it 
is useful to make the variable change $u=w-\epsilon y$. This leads 
to a double integral of the form 
\[ \int \dd w \int_0^\infty \dd y I_{jl}(w,y) \; .\]
Thanks to the fast decay of integrand $I_{jl}(w,y)$ in any direction of 
plane $(w,y)$, the integrals upon $w$ and $y$ can be exchanged. Then 
in the integral upon $w$, we make the variable change $w \rightarrow \eta$ 
with
\[ \eta= (1+2y)^{1/2} \left(w-\frac{y(1+y)}{1+2y} +
\frac{iky}{\epsilon(1+2y)} \right)  \;   .       \]
This provides
\begin{multline}
\label{intAjl}
A_{jl}(k;z)= \frac{\epsilon}{\sqrt{2\pi}}\int_0^\infty \dd y 
\frac{1}{(1+2y)^{1/2}} \\
\times \exp \left( -(z+1)y -\epsilon^2 \frac{y^3(2+y)}{6(1+2y)} 
-\frac{k^2y^2}{2(1+2y)} -\frac{ik\epsilon y^2(1+y)}{1+2y}\right) \\
\times \int \dd \eta \exp(-\frac{\eta^2}{2}) 
\left(\frac{\eta}{1+2y)^{1/2}} + \frac{\epsilon y(1+y)}{1+2y} 
-\frac{iky}{1+2y} \right)^j \\
\times \left(\frac{\eta}{1+2y)^{1/2}} -\frac{\epsilon y^2}{1+2y} 
-\frac{iky}{1+2y} \right)^l 
\end{multline}
The integral upon $\eta$ can be easily performed, thanks to the 
simple dependence of the corresponding integrand with respect to 
$\eta$, namely a Gaussian times a polynomial. The result is a 
combination of algebraic functions of $y$ with coefficients which reduce to positive integer powers of $k$. Thus, the remaining integral upon $y$ 
does converge for any complex value of $k$ and $z$, thanks to 
the presence of factor 
\[ \exp\left( -\epsilon^2 \frac{y^3(2+y)}{6(1+2y)}  \right) \]
which ensures a fast integrable decay of the integrand when $y \to \infty$.
That fast decay guarantees that $A_{jl}(k;z)$ is an entire function of both 
complex variables $k$ and $z$.

\bigskip

Integral representation (\ref{intAjl}) can be specified to $k=0$, 
$j=0,1$ and $l=0,1$. This provides useful expressions 
for functions $A_{00}(z)$, $A_{10}(z)$, $A_{01}(z)$ and $A_{11}(z)$ 
which are analogous to formula (\ref{III23}) for $\Delta(z)$. 
That formula is derived as follows. First, we compute 
$\dd \Delta/\dd z$ from expression (\ref{C9}) specified to $k=0$ 
in terms of functions $A_{jl}(z)$ and of their derivatives 
with respecto $z$. Using the integral 
representations (\ref{C5}), such derivatives are then expressed in terms of 
the  $A_{jl}(z)$'s by combining differentiation under the integral 
sign and integration by parts. This allows us to infer that 
$\Delta(z)$ is the solution of the first order differential equation 
\begin{equation}
\label{equadiffD}
\frac{\dd \Delta}{\dd z} -\Delta=3 \; \epsilon \frac{\dd A_{00}}{\dd z} 
- \epsilon A_{00} \; ,
\end{equation}
with the boundary condition at infinity $\Delta(z) \to 0$ 
when $z=x \to \infty$. A straightforward application of the 
constant-variation method leads to 
\begin{equation}
\label{intDelta}
\Delta(z)=3 \; \epsilon A_{00}(z) -2 \; \epsilon \exp(z) \;\int_z^\infty \dd z'
\exp(-z') A_{00}(z') \; .
\end{equation}
Eventually, we use the above integral representation of $A_{00}(z')$ into 
expression (\ref{intDelta}), and we exchange integrals upon $z'$ and $y$ 
thanks to absolute convergence. Since the dependence in $z'$ 
reduces to  simple exponential factor $\exp (-z'(1+y))$,
the integral upon $z'$ is readily done, and  
this eventually leads to formula (\ref{III23}).


\begin{thebibliography}{99} 

\bibitem{GP86} A. Gervois, J. Piasecki, {\it J.Stat.Phys.}~{\bf 42}:1091-1102 (1986)
\bibitem{JP83} J. Piasecki, { \it J.Stat.Phys.}~{\bf 30}:185 (1983).
\bibitem{JP1986} J. Piasecki, {\it Phys.Lett.A}~{\bf 114}:245-249 (1986)
\bibitem{JPRS2006} J. Piasecki, R. Soto, {\it Physica A}~{\bf 369}:379-386 (2006)
\bibitem{BCG2000} A.V. Bobylev, J.A. Carillo, I.M. Gamba, {\it J.Stat.Phys.}
~{\bf 98}:743-773 (2000)
\bibitem{BC2002} A.V. Bobylev, C. Cercignani, {\it J.Stat.Phys.}~{\bf 106}:1019 (2002)
\bibitem{MP2007} Ph.A. Martin, J. Piasecki, {\it J.Phys.A:Math. Theor.}~{\bf 40}:361-369 (2007)
\bibitem{UFM63} G.E. Uhlenbeck, G.W. Ford, E.W. Montroll, Lectures in statistical physics (American Mathematical Society, 1963)
\bibitem{ED75} M.H. Ernst and J.R. Dorfman, 
{\it J.Stat.Phys.}~{\bf 12}:311-359 (1975)
\bibitem{MHE1984} M.H. Ernst, {\it J.Stat.Phys.}~{\bf 34}:issues 5,6 (1984)
\bibitem{CDT2005} F. Coppex, M. Droz, E. Trizac, {\it Phys.Rev.E}~{\bf 72}: 021105 (2005)

\end{thebibliography}
\end{document}